\documentclass{aastex}
\usepackage{emulateapj5}

\newcommand{\kms}{\hbox{ km\thinspace s$^{-1}$}}    %kms -1

\slugcomment{Submitted to AJ}

\shorttitle{Guerrero et al.}
\shortauthors{Structure of the Owl Nebula}

\begin{document}


\title{Physical Structure of Planetary Nebulae. I. The Owl 
Nebula\altaffilmark{1}}


\author{Mart\'{\i}n A. Guerrero and You-Hua Chu}
\affil{Astronomy Department, University of Illinois at Urbana-Champaign,
    Urbana, IL 61801}
\email{mar@astro.uiuc.edu, chu@astro.uiuc.edu}
\author{Arturo Manchado}
\affil{Instituto de Astrof\'\i sica de Canarias, La Laguna
E-38200, Tenerife, Spain \\
Consejo Superior de Investigaciones Cient\'{\i}ficas, Spain}
\email{amt@ll.iac.es}
\and
\author{Karen B.\ Kwitter\altaffilmark{2}}
\affil{Department of Astronomy, Williams College, Williamstown, MA 01267} 
\email{kkwitter@williams.edu}


\altaffiltext{1}{
Based on observations made with the William Herschell Telescope operated 
on the island of La Palma by the Isaac Newton Group in the Spanish 
Observatorio del Roque de Los Muchachos of the Instituto de 
Astrof\'{\i}sica de Canarias, and with the Burrell Schmidt Telescope of 
the Warner and Swasey Observatory, Case Western Reserve University.
}

\altaffiltext{2}{Visiting Astronomer, Kitt Peak National Observatory, 
    National Optical Astronomy Observatories, which is operated by the 
    Association of Universities for Research in Astronomy, Inc. (AURA) 
    under cooperative agreement with the National Science Foundation.}

\begin{abstract}
The Owl Nebula is a triple-shell planetary nebula with the outermost
shell being a faint bow-shaped halo.  
We have obtained deep narrow-band images and high-dispersion echelle
spectra in the H$\alpha$, [O~{\sc iii}], and [N~{\sc ii}] emission 
lines to determine the physical structure of each shell in the nebula.
These spatio-kinematic data allow us to rule out hydrodynamic models 
that can reproduce only the nebular morphology.
Our analysis shows that the inner shell of the main nebula is slightly
elongated with a bipolar cavity along its major axis, the outer nebula
is a filled envelope co-expanding with the inner shell at 40 \kms,
and the halo has been braked by the interstellar medium as the Owl 
Nebula moves through it.
To explain the morphology and kinematics of the Owl Nebula, 
we suggest the following scenario for its formation and evolution.
The early mass loss at the TP-AGB phase forms the halo, and the
superwind at the end of the AGB phase forms the main nebula.
The subsequent fast stellar wind compressed the superwind to form
the inner shell and excavated an elongated cavity at the center,
but has ceased in the past.
At the current old age, the inner shell is backfilling the central
cavity.

\end{abstract}



\section{Introduction}

Planetary nebulae (PNs) consist of stellar material ejected by
low- and intermediate-mass stars at late evolutionary stages. 
As these stars lose mass via dense slow winds at the asymptotic
giant branch (AGB) phase followed by tenuous fast winds,
the dynamic interactions between the fast and slow winds produce
the shell structure of PNs.
A wide variety of morphologies has been observed in PNs.
These morphologies can be successfully reproduced by hydrodynamical
models of interacting stellar winds; however, a nebular morphology 
can often be reproduced similarly well by several models with different 
physical structures and geometries, indicating that morphology alone
cannot constrain the models adequately.
Spatio-kinematic data provide additional constraints to models,
but only few PNs have such observations for critical comparisons.
Therefore, we have started a program to analyze high-resolution 
images and high-dispersion spectra of PNs in multiple nebular 
lines, in order to determine the physical structure of PNs.
By comparing the results with hydrodynamical models, we hope to 
understand better the formation and evolution of PNs.

The first PN we have analyzed is NGC\,3587, better known as the ``Owl 
Nebula" because of its morphological resemblance to an owl's face.
The large angular size ($> 3\arcmin$) and symmetrical morphology 
of the Owl Nebula allow us to analyze its physical structure in detail
even using ground-based imaging and spectroscopic observations.
The Owl Nebula is a triple-shell PN, consisting of a round double-shell 
main nebula and a faint bow-shaped outer halo \citep{cja87,kcd92}.  
The morphology of the halo suggests an interaction with the surrounding 
interstellar medium (ISM).
Thus the Owl Nebula provides a rare opportunity for us to probe the
extended mass loss history and the dissipation of a PN into the ISM.
However, because of its low surface brightness, previously there was 
only one spatio-kinematic study of the Owl Nebula \citep{sa85} and 
it was limited to only the bright inner shell of the main nebula.  
We have obtained long-slit, high-dispersion spectroscopic observations 
and narrow-band images of the Owl Nebula in the H$\alpha$, [O~{\sc iii}] 
$\lambda$5007, and [N~{\sc ii}] $\lambda$6583 emission lines.  
These data have been used to construct a spatio-kinematic model 
of the main nebula, and to gain insights into the structure 
and nature of the halo.  
The results of our analyses are reported in this paper.

\section{Observations}

Narrow-band images of the Owl Nebula in the H$\alpha$, [O~{\sc iii}] 
$\lambda$5007, and [N~{\sc ii}] $\lambda$6583 emission lines were 
obtained in 1991 June with the 0.6\,m Burrell Schmidt Telescope at Kitt 
Peak National Observatory (KPNO). 
The central wavelengths and FWHMs of the interference filters are 
6558 \AA\ and 20 \AA\ for H$\alpha$, 
4998 \AA\ and 23 \AA\ for [O~{\sc iii}], and 
6582 \AA\ and 21 \AA\ for [N~{\sc ii}].  
The transmission of the H$\alpha$ filter at 6583 \AA\ is $\sim 25\%$
of that at 6563 \AA, while the transmission of the [N~{\sc ii}] filter
at 6563 \AA\ is $\sim 6\%$ of that at 6583 \AA. 
With a [N~{\sc ii}] $\lambda$6583 to H$\alpha$ line ratio of $\sim 0.5$ 
\citep{kh01}, we expect the H$\alpha$ image to be contaminated by
the [N~{\sc ii}] line at $\sim 12\%$ level, and the [N~{\sc ii}] 
image by the H$\alpha$ line at a similar level.

The detector was the ST2K 2048$\times$2048 CCD (also known as S2KA), but 
only the central 1200$\times$1200 pixels were read out. 
% to avoid the vignetted outer region.  
The pixel size is 21 $\mu$m, corresponding to a spatial scale of 
$2\farcs07$ pixel$^{-1}$ and providing a field of view of $41\farcm4 
\times 41\farcm4$.  
The angular resolution was $\sim3\farcs2$, as determined from the 
FWHM of field stars in the images. 
The exposure time was 20 minutes for each image.  
Figure~1 shows the H$\alpha$, [O~{\sc{iii}}], and [N {\sc{ii}}] 
images of the Owl Nebula.  
A [N {\sc{ii}}]/[O {\sc{iii}}] ratio map is shown in Figure~2.  
To illustrate the brightness variations quantitatively, the 
[O{\sc{iii}}] and [N~{\sc{ii}}] surface brightness profiles along 
PA $-45\arcdeg$ and $45\arcdeg$ are presented in Figure~3.  

Long-slit echelle observations of the Owl Nebula were obtained using the 
Utrecht Echelle Spectrograph (UES) on the 4.2~m William Herschell 
Telescope (WHT) at the Observatory of Roque de los Muchachos (La Palma, 
Spain) on 1993 June 5 and 6.   
The spectrograph was used in the long-slit mode to obtain single-order 
observations for an unvignetted slit length of {160\arcsec}.  
Broad (FWHM = 100 \AA) filters centered at 6563 \AA\ and 5007 \AA\ 
were used to isolate the H$\alpha$ + [N~{\sc ii}] $\lambda\lambda$6548,6583 
lines and the [O~{\sc iii}] $\lambda$5007 line for their respective 
observations.
The 79 line~mm$^{-1}$ echelle grating and the long-focus camera were 
used; the resultant reciprocal dispersion was 2.84 \AA~mm$^{-1}$ at 
H$\alpha$ $\lambda$6563, and 2.14 \AA~mm$^{-1}$ at [O~{\sc iii}] 
$\lambda$5007.  
% 2.144 and 2.837
The data were recorded with a 1024$\times$1024 Tektronix CCD with a pixel 
size of 24 $\mu$m.  
This configuration provides a spatial scale of $0\farcs36$ pixel$^{-1}$ 
and a sampling of 3.1 km~s$^{-1}$~pixel$^{-1}$ along the dispersion 
direction.  
The slit width was $1\farcs1$, and the resultant instrumental FWHM 
was 6.6\kms.  
%Wavelength calibration was better than 0.2 pixels (0.6\kms).
The angular resolution, determined by the seeing, was $\sim0\farcs9$ 
FWHM. 

For the H$\alpha$ observations, we placed the slits along the ``owl's eyes''
at PA $-45\arcdeg$ and along the orthogonal direction at PA $45\arcdeg$,
as marked on the H$\alpha$ images in Fig.\ 1,  because these directions 
define the most prominent axes of symmetry in the nebula.  
The large angular size of the Owl Nebula ($> 3\arcmin$) required 
multiple slit positions to cover the nebula along these two PAs. 
The [O~{\sc iii}] observation was made at only one slit position 
along PA $45\arcdeg$, covering only the main nebula and a small section of 
the halo.  
The integration times, central wavelengths, position angles, and 
offsets from the central star of the long-slit echelle observations 
are given in Table~1.  
Grey-scale plots of the echellograms are presented in Figure~4.

\section{Results}

\subsection{Morphology}

The images of the Owl Nebula (Fig.~1) clearly show a triple-shell 
structure;  these shells will be referred to as the inner shell, the 
outer shell, and the halo.  
The detailed morphology and the variations of excitation of these 
shells are described below.  
% The morphology of the inner and outer shells has been described 
% by \citet{cp00} based on images in a larger set of nebular emission 
% lines.  
% Similarly, the morphology of the outer halo has been described by 
% \citet{kcd92} and \citet{ha97}.  

The inner shell is roughly confined by an ellipse, $182\arcsec \times 
168\arcsec$ in size, with its major axis oriented along the NW-SE 
direction (Fig.~1-{\it center-left}, and {\it bottom-left}).  
The main morphological feature of the inner shell is the ``face of an 
owl'', which is best seen in the H$\alpha$ and [O~{\sc iii}] images.
The ``eyes of the owl" are two irregularly shaped regions of diminished 
surface brightness aligned along the major axis of the inner shell,
with each eye $\sim35\arcsec$ in size.
The ``forehead'' and ``beak'' are two regions of enhanced surface 
brightness separated by $\simeq1\arcmin$ along the minor axis of the 
inner shell.
In the [N~{\sc ii}] image, the owl's eyes appear larger and the 
brightness enhancement in the owl's forehead and beak is not as 
prominent as in the H$\alpha$ and [O~{\sc iii}] images (see Fig.~3).  
The [N~{\sc ii}] image also shows differences from the [O~{\sc iii}]
image in the transition region between the inner and outer shells. 
The outer edge of the inner shell is regular and smoothly merges 
into the outer shell in the [O~{\sc iii}] image, but appears 
irregular and shows a noticeable break in surface brightness
in the [N~{\sc ii}] image.  
% the [O~{\sc iii}] emission being slightly less extended.  

The outer shell is almost circular.  
With a diameter of {218\arcsec}, the outer shell is only 20--30\% larger 
than the inner shell, making it one of the thinnest (relative to the 
inner shell) outer shell among multiple-shell PNs \citep{cja87}.  
The surface brightness of the outer shell decreases outwards in the 
H$\alpha$ and [O~{\sc iii}] images, but shows pronounced 
limb-brightening in the [N~{\sc ii}] image along the PAs from 
{$-$15\arcdeg} to {$+$50\arcdeg}, and from {180\arcdeg} to {230\arcdeg}.

The halo is more prominent in the [O~{\sc iii}] image than 
in the H$\alpha$ or [N~{\sc ii}] images (Fig.~1-{\it right}), as 
previously noted by \citet{ha97}.  
This behavior is commonly seen in haloes of PNs and is most likely an 
effect of the hardening of the ionizing radiation \citep{ppm83,gm99}.
At the faintest level, the 3 $\sigma$ contour of the halo 
in Fig.~1-{\it right-center} is remarkably circular with a  
diameter of 350\arcsec, although the overall morphology of the 
halo is asymmetrical.
The northeast part of the halo is brighter and less extended 
with a noticeable limb-brightening from PA$\sim-10\arcdeg$ to 
PA$\sim+125\arcdeg$, while in the opposite direction the halo is 
fainter and more extended with a gradual surface brightness drop off.
The bright main nebula appears displaced in the halo towards its
bright northeast rim.
This morphology suggests a bow-shock formed by the motion of the Owl 
Nebula through the ISM, as discussed later in \S4.2.
% the emission 
%detected beyond the edge-brightened northeast rim indicates that
%the apex of the bow-shock is tilted against the sky plane.

%It is worth noting that some of the morphological features in the
%main nebula are related to those in the halo
%For example, bright [N~{\sc ii}] emission is seen in the outer shell 
%at PA$\sim${18\arcdeg}, while an [O~{\sc iii}]-dim region in the 
%halo is observed at this PA.  
%As shown in Figure~2, the highest [N~{\sc ii}]/[O~{\sc iii}] ratio 
%is found in a relatively compact region at 80\arcsec--95\arcsec\ 
%from the central star at PA$\sim${18\arcdeg}.  

It is worth noting that some excitation features in the main nebula 
are correlated with those in the halo.
As shown in the [N~{\sc ii}]/[O~{\sc iii}] ratio map of the main nebula
(Fig.~2), the highest [N~{\sc ii}]/[O~{\sc iii}] ratio is found in 
the northeast quadrant, peaking particularly in a compact region 
at 80\arcsec--95\arcsec\ from the central star at PA$\sim${18\arcdeg}.  
In this compact region, the [N~{\sc ii}] emission, as well as the
H$\alpha$, is enhanced.
In the halo behind this compact region, the [O~{\sc iii}] image shows 
a conical ``shadow'' along a line that can be traced back to the 
central star (Fig.~1-{\it center-right}).            
%At the same PA in the halo, 
%The [O~{\sc iii}] shadow is also present in the halo, where the rim 
%observed in [O~{\sc iii}] fades at PA$\sim+18\arcdeg$ \citep{ha97} 
%and the faint [N~{\sc ii}] emission is enhanced.  
A similar feature is observed in the halo of NGC\,6853, the Dumbbell 
Nebula \citep{m96}. 
This phenomenon is most likely caused by a dense low-ionization knot 
within the inner shell of the nebula which shields the halo from the 
light of the PN nucleus.  
%
%Elliptical halo: $300\arcsec\times315\arcsec$, PA$\sim50\arcdeg$, 
%displacement $\sim23\arcsec$. \\ 
%(+21,-10) 

\subsection{Kinematics}

The echellograms of the H$\alpha$, [O~{\sc iii}], and [N~{\sc ii}] 
lines in Fig.~4 show position-velocity ellipses that are 
characteristic of expanding shells.
The systemic velocity of the Owl Nebula measured from the centroid 
of the position-velocity ellipses is $\simeq 12\kms$ relative to
the local standard of rest.  [All radial velocities in this paper 
are $V_{\rm LSR}$.]
The velocity split at the center of the nebula is 55.4\kms\ in 
H$\alpha$, 73.3\kms\ in [O~{\sc iii}], and 78.9\kms\ in [N~{\sc ii}]
%$78.9\pm1.2\kms$ and the radial velocity in the 
These measurements are consistent with the systemic velocity and 
expansion velocities reported by \citet{sa85}.
%
% 1985, Sabbadin 
%      VRAD = 6 (3) km/s   VRAD = 8 (5)  km/s  (1918, Campbell & Moore)
%                          VRAD = 3 (15) km/s  (1962, Chopinet)
%                          VRAD = 5 (4)  km/s  (1976, Acker)
%      Vexp[O III] = 28 (1.5)   km/s     ours: 36.7
%      Vexp[H I]   = 27.5 (1.5) km/s
%      Vexp[N II]  = 40 (1)     km/s     ours: 39.5
%

None of the lines show obvious tilt along the major- or minor-axis of 
the inner shell.
A line tilt is indicative of an aspherical shell.
The lack of obvious line tilt, therefore, implies that either the 
inner shell of the Owl Nebula is nearly spherical as shown in the 
direct images or its axis of symmetry is close to the line of 
sight.  
In the latter case, the nearly circular projection of the inner 
shell places a stringent constraint on the axis of symmetry: 
the longer the axis of symmetry, the closer it must be to the 
line of sight.
It is very unlikely that the Owl Nebula has its long axis pointing 
at us exactly.  
Therefore, we suggest that the Owl Nebula has only a small deviation 
from a spherical shell.  
% is nearly spherical with a deviation from a spherical shell less 
% than $\sim$10\%. 

The thickness of an expanding shell can be estimated from the
brightness variations of an echelle line image along the systemic 
velocity, as this iso-velocity slice samples the nebular material 
whose radial expansion is perpendicular to the line of sight.  
Along the minor axis of the Owl Nebula, the H$\alpha$ and 
[O~{\sc iii}] lines show a thick shell with a small central
cavity $\sim$30\arcsec\ in size, while the [N~{\sc ii}] line
shows a thin shell with a large central cavity 
$80\arcsec-90\arcsec$ in size.  
The inner shell shows pronounced limb-brightening in the 
[N~{\sc ii}] line, in sharp contrast to the uniform 
appearance in the H$\alpha$ line.
The [O~{\sc iii}] line has an overall line morphology similar to 
that of H$\alpha$, rather than [N~{\sc ii}], but the smaller 
thermal broadening of the [O~{\sc iii}] line makes it possible 
to resolve brightness variations.  
Two patches of brighter emission at the systemic velocity are 
detected in the inner shell at positions roughly coincident with 
the forehead and beak of the owl's face.  
The [O~{\sc iii}] line of the inner shell also shows a 
position-velocity ellipse of brighter emission with radius 
70\arcsec\ which corresponds to a change in the slope observed 
in the [O~{\sc iii}] brightness profile (Fig.~3).

The echellograms of the H$\alpha$ and [N~{\sc ii}] lines along the 
major axis (PA $-$45\arcdeg) show large brightness variations along 
the position-velocity ellipses, especially on the receding half of 
the shell.  
While the [N~{\sc ii}] line shows a large central cavity similar to 
that along the minor axis, the H$\alpha$ line shows a larger and 
more complex central cavity than along the minor axis.  
The central cavity in the H$\alpha$ line appears bipolar with each 
lobe about 60\arcsec--70\arcsec\ in extent;  
the NW lobe is displaced to the red side and the SE lobe to the 
blue side of the systemic velocity.

The outer shell is detected in the echellograms in Fig.~4.  
The [N~{\sc ii}] line shows that the position-velocity ellipse 
of the outer shell merges together with that of the inner shell 
near the projected center of the nebula, indicating that the 
inner and outer shells have similar expansion velocities.   
The H$\alpha$ and [O~{\sc iii}] lines show a filled featureless 
outer shell, but the [N~{\sc ii}] line shows enhanced emission 
along the outer edge at the NE end of the minor axis and the SE 
end of the major axis.

The halo is detected in the [N~{\sc ii}] and [O~{\sc iii}] lines, 
but not in the H$\alpha$ line (Fig.~4) because the larger thermal 
width of hydrogen spreads its emission over many pixels and lowers 
the S/N excessively.   
Only the inner 26\arcsec\ of the halo toward the northeast is 
detected, corresponding to the brightest part of the halo
(within the 10$\sigma$ contour in Fig.~1-{\it center-left}).
No line splitting is observed in the halo.  
The intrinsic FWHM (removing the instrumental contribution to the 
observed width) of the [N~{\sc ii}] and [O~{\sc iii}] lines are 
$\sim$12\kms\ and $\sim$25\kms, respectively.  
The large difference in the intrinsic FWHM of these two lines 
indicates that the [N~{\sc ii}] and [O~{\sc iii}] emission 
originates from regions with different physical conditions.  
Despite the difference in FWHM, the [N~{\sc ii}] and [O~{\sc iii}] 
lines show similar variations in centroid velocity with position:    
the velocity varies from $\sim$12\kms\ (the systemic velocity of 
the Owl Nebula) outwards to $\sim$22\kms\ at the bright rim of the 
halo.   
Similar velocity variations in the halo have been reported in 
NGC\,6751, and they are caused by interstellar braking 
\citep{Cetal91}.

\section{Discussion}

\subsection{The Spatio-Kinematic Structure of the Main Nebula}

To build a spatio-kinematic model of the Owl Nebula, we 
adopt the methodology used by \citet{Getal00} for the analysis of 
NGC\,6891 and make three assumptions: 
(1) material within the nebula moves exclusively along radial 
    directions,  
(2) each shell was ejected at different times, and 
(3) each shell is composed of subshells that expand homologously, 
    i.e., the velocity increases linearly with distance from the 
    center of the shell.  
Following these assumptions, the elliptical morphology and the 
position-velocity ellipse of the inner shell indicate that the 
boundary of this shell is ellipsoidal in shape.  
The best-fit ellipsoid has a radius of 93\arcsec\ along the polar 
direction and a radius of 83\arcsec\ in the equatorial plane, and
its real major axis is tilted by 10\arcdeg--30\arcdeg\ with respect 
to the line of sight toward the SE along the apparent major axis.
The polar and equatorial expansion velocities are $\sim$45\kms\ 
and $\sim$40\kms, respectively.

The H$\alpha$ and [O~{\sc iii}] echellograms show that the inner 
shell is partially filled.  
Moreover, the differences between the H$\alpha$ echellograms 
along the major and minor axes indicate that the thickness of 
the inner shell is non-uniform and depends on the polar angle,
as suggested previously by \citet{pr72}.  
The distribution of the emission in the H$\alpha$ echellogram along 
the major axis is best explained by the inclusion of a pair of
bipolar cavities along the same axis of symmetry as the inner 
shell \citep{sa85}.
Our integrated model for the inner shell of the Owl Nebula is 
illustrated in Figure~5-{\it left}.  
If the emitting material in the inner shell has a uniform density,
the surface brightness of the inner shell will have the morphology 
shown in Fig.~5-{\it right}, which nicely reproduces the owl's 
face morphology exhibited in the H$\alpha$ and [O~{\sc iii}] images
in Fig.~1.
The [N~{\sc ii}] morphology of the inner shell is different, being 
thinner and displaced toward the outer edge of the H$\alpha$ and 
[O~{\sc iii}] emission.
These differences can be explained by ionization stratification,
with the low-ionization species distributed in a thin shell exterior 
to the high-ionization species.

The observed morphology and kinematics of the outer shell are 
consistent with a spherical shell expanding at {40\kms}.
The lower surface brightness of the outer shell is most likely
caused by a lower density.
There is no evidence of an asymmetric geometry for the outer shell; 
therefore, we simply conclude that the outer shell of the Owl Nebula 
is a low-density shell around the denser inner shell, as depicted 
in Fig.~5-{\it left}.
%, but the small fraction of 
%this shell that can be observed does not allow us to rule out an 
%asymmetric shell.  

Several previous models of the Owl Nebula are available, but they 
cannot explain the morphology and the kinematics simultaneously.
In the framework of interacting stellar winds, hydrodynamical 
calculations have been used by \citet{ibf92} (see their Figure 9)
and \citet{fm94} (see their Figures 6 and 7) to reproduce the 
owl's face morphology.  
In both cases, the best models consist of prolate shells with 
large axial ratios ($\sim 2$) and inclination angles 
($\ge 40\arcdeg$) against the line of sight.  
The nebular kinematics expected from these models will produce
marked tilt in the velocity-position diagram.
Our echelle observations (Fig.~4) contradict these predictions
directly, ruling out these models.
Alternatively, Zhang, \& Kwok (1998) used a complex geometric
model with artificial density distribution to synthesize the
Owl's morphology.
Their model reproduced the Owl's eyes, but also produced 
extended ``ears" along the major axis (see their Figure 4),
which are not observed in direct images; thus we consider this 
model also unsatisfactory.
{\it Viable models of PNs must explain nebular kinematics as 
well as nebular morphology.}

\subsection{The Spatio-Kinematic Structure of the Halo}

As seen in Fig.~1-{\it center-right}, the morphology of the 
Owl Nebula's halo suggests that the nebula moves through the 
ISM toward the northeast direction.  
The direction of this motion is in agreement with the proper 
motion of its central star along PA $30\arcdeg$$\pm20\arcdeg$ 
\citep{C74}.  
The ram pressure of the ISM compresses the leading edge of the
halo producing the bow-shaped bright northeast rim.
Furthermore, the deceleration of the halo material on the
leading edge explains the displacement of the main nebula in
the halo.
Note, however, that faint emission exists exterior to the leading 
edge of the halo defined by the 10$\sigma$ contour, and that the
3$\sigma$ contour is remarkably round.
This can be easily reproduced if the direction of the translational
motion of the Owl Nebula through the ISM is tilted at a large angle 
against the sky plane as sketched in Figure~6-{\it left}.
If the apex of the nebular motion is closer to the line of sight than 
the sky plane, the projection of the aspherical outer edge of the
halo (3$\sigma$ contour) can appear round, and the faint halo off 
the leading edge can be projected beyond the bright rim, contributing
to the emission between the 3$\sigma$ and 10$\sigma$ contours to the
northeast.
We have synthesized the halo morphology, assuming linear azimuthal
variations in emissivity and thickness of the halo between the
leading and trailing sides to account for the compression of material 
at its leading edge.  
Our model consists of an ellipsoidal shell with major and minor axes 
400\arcsec\ and 360\arcsec, respectively, tilted 30\arcdeg\ with the 
line of sight.  
For a fractional shell thickness of 10\% on the leading edge, and an 
emissivity contrast of 9 and a thickness contrast of 3 between the 
leading and trailing edges of the halo, the expected halo morphology 
shown in Figure~6-{\it right} agrees reasonably well with the 
observed halo.

The kinematics of the halo also shows evidence of its interaction
with the ambient ISM.
The echellograms in Fig.~4 detect the brightest parts of the halo,
i.e., within the 10$\sigma$ contour in Fig.~1~{\it center-right}.
The line profiles of the halo are narrow without any line splitting.
The velocity of the halo varies from $\sim$12 \kms, the systemic 
velocity of the main nebula, outwards to $\sim$22 \kms.
Such velocity variations are typical for interstellar braking, as
observed in the PN NGC\,6751 \citep{Cetal91} and the LBV nebula 
S119 \citep{DC01}, where the nebular material is decelerated from
the systemic velocity of the nebula to that of the ambient ISM.

%The confinement of the halo of the Owl Nebula by the ISM is puzzling.
%The high Galactic latitude ($b = 57.0\arcdeg$) of the Owl Nebula 
%places it at a large height above the Galactic plane where the 
%density of the ISM is expected to be low.
%The distance to the Owl Nebula has been estimated to be between
%0.4 kpc and 1.4 kpc \citep{p84,cks92,n99}, and hence its height
%above the Galactic plane may be between 350 pc and 1,150 pc. 
%Even with the short distance estimate, the density of the ISM is 
%expected to be $\simeq 0.06$ cm$^{-3}$ \citep{s78,mb81}
% 2 x exp{-z/100}
%}.
%to provide the required thermal pressure ($N_{\rm ISM} \times T \sim 
%10^3$ K~cm$^{-3}$) to confine its halo \citep{b92}.  
% The contribution of the ISM material to the total mass of the halo is 
% also expected to be small, $\le 0.002 M_\odot$.  
%Alternatively, the halo of the Owl Nebula is interacting with material 
%previously ejected by the star during the AGB phase.  

\subsection{The Evolutionary Status of the Owl Nebula}

We will determine the kinematical age of each shell of the Owl 
Nebula to assess its evolutionary status.
Note that the relationship between the kinematical age and the 
evolutionary age of each shell of a PN may not be straightforward 
\citep{vmg02}. 
Furthermore, the kinematical ages are plagued by the uncertainty
in the distance.
For the Owl Nebula, the distance has been estimated to be between 
0.4 kpc and 1.4 kpc \citep{p84,cks92,n99}, which implies a factor 
of 3 uncertainty in the kinematical ages.
Thus, our discussion below provides only a qualitative view of the
nebular evolution.

The kinematical age of the main nebula can be calculated from the 
expansion velocity and size of the outer shell \citep{s97}.  
For an expansion velocity of {40\kms} and an angular size of 
{218\arcsec}, the kinematical age is $12,900 \times d$ yr, where 
$d$ is the distance in kpc.  
This age can be combined with the ionized mass of the nebula to derive
an average mass loss rate.
As given by Pottasch (1984), the ionized mass of a PN is 
\begin{equation}
M_{\rm i} [{\rm M}_\odot] = 2.2 \times F({\rm H}\beta)^{0.5} T_{\rm e}^{0.44} 
          \epsilon^{0.5} \theta^{1.5} d^{2.5}
%  334.0 \times F({\rm H}\beta) d^2 T_{\rm e}^{0.88} N_{\rm e}^{-1}
\end{equation}
where $\epsilon$ is the filling factor, and $\theta$ is the angular 
radius in arcsec.  
Adopting $\epsilon = 0.5$, an observed H$\beta$ flux $F({\rm H}\beta)$
= $3.2{\times}10^{-11}$ erg cm$^{-2}$ s$^{-1}$ \citep{k76}, and 
electron temperature $T_{\rm e}$ = $11,000$ K \citep{ksk90}, we 
find an ionized mass of $0.60 \times d^{2.5} {\rm M}_\odot$.  
Thus, the average mass loss rate is $5 \times 10^{-5} \times d^{1.5} 
{\rm M}_\odot$ yr$^{-1}$, which is a lower limit for the copious
mass loss at the AGB phase. 
% rms density ~ 100

The halo of the Owl Nebula has no measurable expansion.
It is not clear whether it has been decelerated by the 
ISM or it was ejected at small velocities.
Assuming an expansion velocity of 5 \kms, the {350\arcsec} diameter
of the halo implies an age of $1.5 \times 10^5 \times d$ yr.
We derive a halo mass of $0.15 \times d^{2.5} {\rm M}_\odot$, and
an average mass loss rate for the halo of 10$^{-6} \times d^{1.5}
{\rm M}_\odot$ yr$^{-1}$.

% According to our H$\alpha$ image, the surface brightness of the halo is 
% $\lq 60$ times the surface brightness of the main shell.  
% Since the halo extends over an area $\sim 1.6$ times larger than the 
% main nebula, the H$\alpha$ flux is $\sim 40$ times smaller than that 
% of the main nebula.  
% rms density ~ 10

The formation of the Owl Nebula can be schematically described as 
the following.
Its halo was formed as the result of an early AGB wind that snow-plowed 
the surrounding ISM.
This AGB wind may be associated with the TP-AGB phase.  
During later stages of the AGB evolution, the central star underwent 
an episode of high mass loss (the ``superwind''), forming the main nebula.  
At its present old age, the Owl Nebula has a large size ($\ge 0.4$ pc)
and a low density ($<$ 100 cm$^{-3}$ in the main nebula), and the
central star has already reached the turning point of its post-AGB track 
towards low luminosities \citep{n99}.  
The diminishing of the ionizing flux causes the recombination of material
within the main nebula and the formation of ionization shadows extending 
into the halo.

Two remarkable features of the Owl's main nebula, namely, 
the presence of a central bipolar cavity and the lack of a bright 
rim, may have significant implications for its formation and 
evolution.
The bipolar cavities in the nebula were probably excavated by the
fast stellar wind of the central star.
However, no bright rim is seen in the inner shell as expected from 
the interaction between the fast wind and the superwind 
\citep{kpf78,fbr90}; furthermore, no measurable fast wind from the 
Owl's central star or diffuse X-ray emission from shocked fast 
wind has been detected \citep{pp91,CGC98}.
We therefore suggest that a fast stellar wind excavated an elongated 
cavity in the nebula, similar to that seen in NGC\,6891 \citep{Getal00},
but the fast wind has subsequently ceased.
Without being pressurized by the fast stellar wind, the nebular
material backfills the central cavity and smears out the bright rim.
The high density along the equatorial plane that was responsible
for shaping the elongated geometry of the central cavity is also 
responsible for producing the forehead and beak of the Owl Nebula 
because of the higher density in the backfilled material.
It ought to be noted that the morphology of the Owl Nebula is 
not unique, as other PNs show similar morphologies, e.g., A\,33 
\citep{hk99}, A\,50 and NGC\,2242 \citep{m96}.  
It is thus very likely that these PNs have all followed similar 
evolutionary paths.  

\section{Summary}

\begin{enumerate} 

\item 
The Owl Nebula is a triple-shell PN, with a slightly elongated inner
shell, a round outer shell, and a bow-shaped halo.
The echelle images of the H$\alpha$ and [O~{\sc iii}] lines show a
thick inner shell with no obvious line tilt; furthermore, the intensity
variations suggest a bipolar central cavity.
The echelle images of the [N~{\sc ii}] $\lambda$6583 line show a thin
shell on the outer edge of the  H$\alpha$ and [O~{\sc iii}] shell, also
without line tilt.
These results rule out the hydrodynamic models which consist of an 
elongated ellipsoidal or bipolar inner nebula with its polar axis tilted
against the line of sight to produce the Owl morphology.

\item 
The halo of the Owl Nebula shows no measurable expansion.
Its velocity varies from $V_{\rm LSR}$ = 12 \kms, the systemic velocity
of the Owl Nebula, at the inner edge to 22 \kms\ at the outer edge of
the bright rim.
The morphology and velocity structure of the halo are caused by the
translational motion of the Owl Nebula through the ISM, a phenomenon
called ``interstellar braking."
The detailed morphology of the halo requires the Owl Nebula to be
moving in the direction toward the northeast, but closer to the line 
of sight than the sky plane.

\item 
The Owl Nebula consists of the AGB wind, with the halo formed
at an early TP-AGB stage, and the main nebula formed during a
period of high mass loss rate near the end of the AGB phase.
The subsequent fast stellar wind from the central star excavated
a bipolar central cavity, but has ceased.  
The backfilling of nebular material has partially erased the 
dynamical imprints of the fast stellar wind.
\end{enumerate}

\acknowledgments

M.A.G.\ and Y.-H.C.\ acknowledge support from NASA Grant NAG5-12255.  
A.M.\ acknowledges support from Grant AYA2001-1658.  
K.B.K.\ and Y.-H.C.\ acknowledge support from NASA Grant NAG5-1888. 
K.B.K.\ also thanks The Research Corporation and the Bronfman Science 
Center at Williams College for support. 
The authors are grateful to KPNO for observing time and support, and 
to the IRAF group for software support.

%% Generally speaking, only the figure captions, and not the figures
%% themselves, are included in electronic manuscript submissions.
%% Use \figcaption to format your figure captions. They should begin on a
%% new page.

%\end{document}

\clearpage

%% No more than seven \figcaption commands are allowed per page,
%% so if you have more than seven captions, insert a \clearpage

%% There must be a \figcaption command for each legend. Key the text of the
%% legend and the optional \label in curly braces. If you wish, you may
%% include the name of the corresponding figure file in square brackets.
%% The label is for identification purposes only. It will not insert the
%% figures themselves into the document.
%% If you want to include your art in the paper, use \plotone.
%% Refer to the on-line documentation for details.

\begin{figure}
\epsscale{0.90}
\plotone{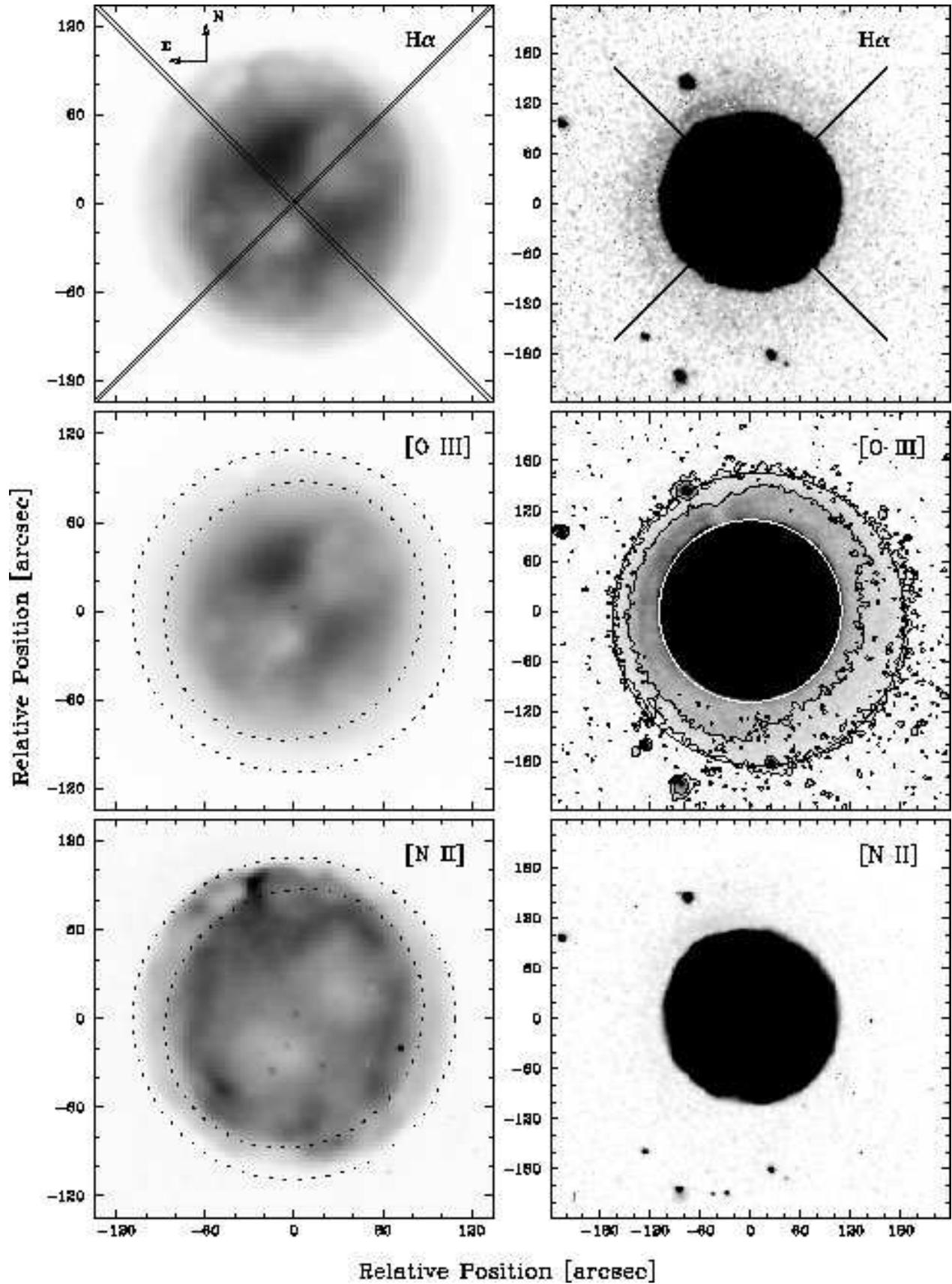}
\caption{
Images of the Owl Nebula in the H$\alpha$ ({\it top}), [O~{\sc iii}] 
$\lambda$5007 ({\it middle}), and [N~{\sc ii}] $\lambda$6583 ({\it 
bottom}) emission lines.  
Each image is displayed at two intensity contrasts and spatial scales 
to highlight the main nebula ({\it left}) and faint halo ({\it right}).  
The slit positions of the echelle observations are plotted over the 
H$\alpha$ images, as explained in the text.  
The boundaries of the inner and outer shells are marked by the dashed 
ellipses overplotted on the [O~{\sc iii}] and [N~{\sc ii}] images of 
the main nebula ({\it middle-left}, and {\it bottom-left}).  
The [O~{\sc iii}] image of the halo ({\it middle-right}) is overlaid 
by the 10 $\sigma$ and 3 $\sigma$ contours of the [O~{\sc iii}] 
emission, as well as two circles marking the boundaries of the outer 
shell and halo.  
\label{images}}
\end{figure}

\begin{figure}
\epsscale{0.5}
\plotone{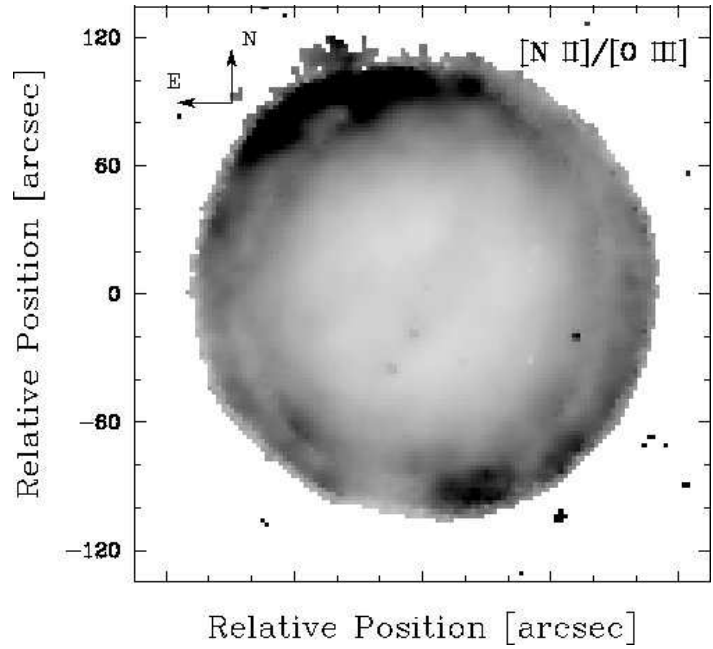}
\caption[]{
[N~{\sc ii}] $\lambda$6583 to [O~{\sc iii}] $\lambda$5007 ratio map of 
the Owl Nebula. 
Dark regions correspond to bright [N~{\sc ii}] emission. 
\label{ratio_map}}
\end{figure}

\begin{figure}
\epsscale{0.5}
\plotone{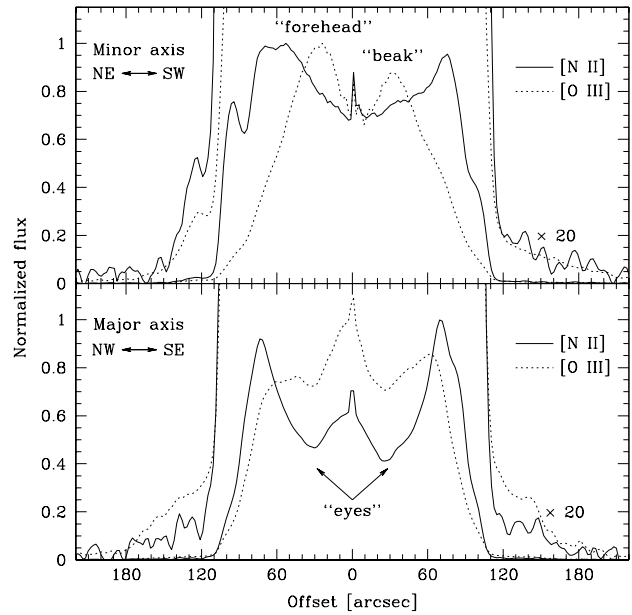}
\caption{Surface brightness profiles of the [O~{\sc iii}] and
[N~{\sc ii}] lines along the minor and major axes of the Owl Nebula.  
The positions of the ``eyes'', ``forehead'' and ``beak'' of the Owl's 
face are marked on these profiles.  
To show the faint halo, these profiles scaled up by a factor of 20 are 
also plotted.  
}
\end{figure}

\begin{figure}
\epsscale{1.0}
\plotone{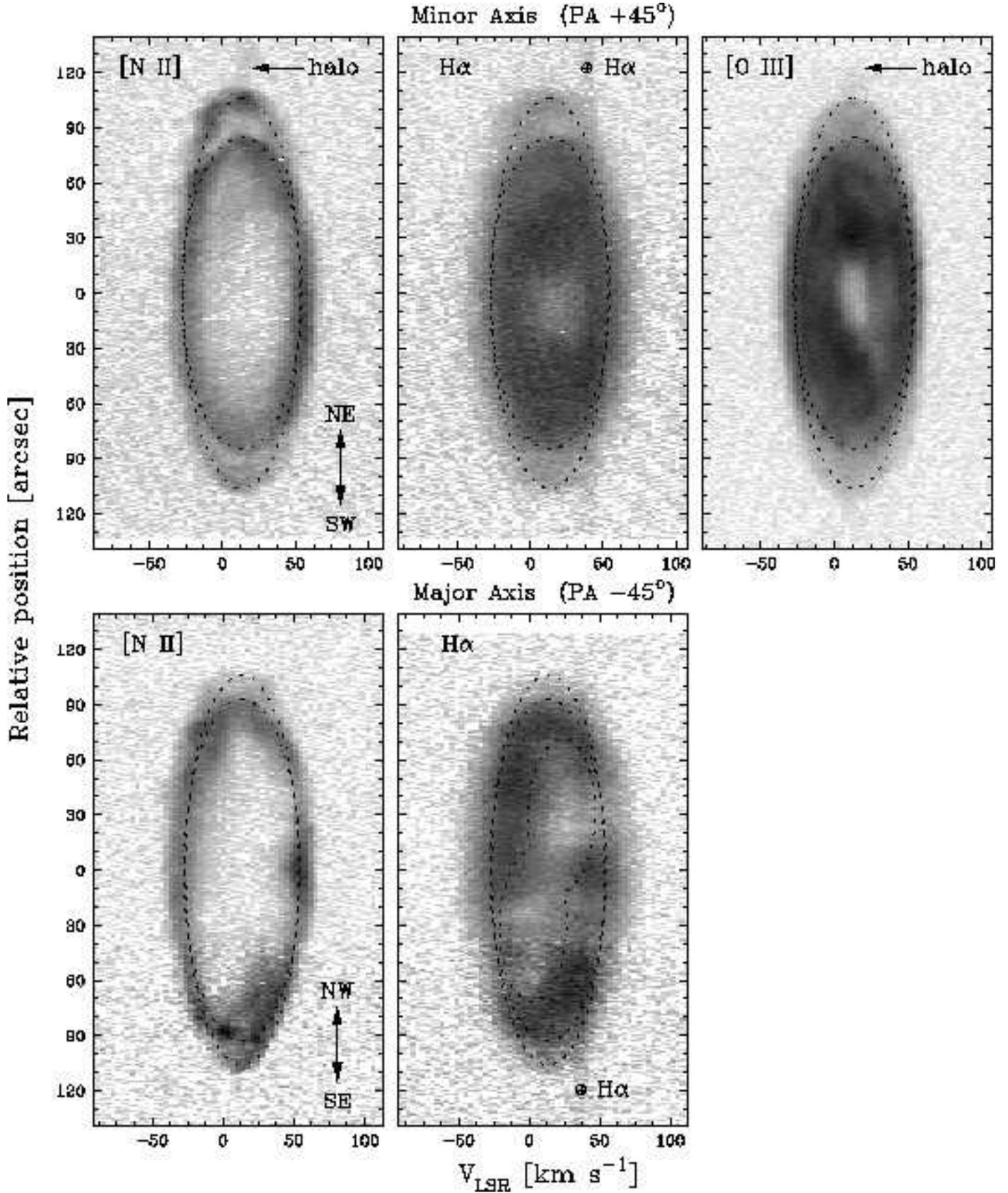}
\caption{
H$\alpha$, [N~{\sc ii}] $\lambda$6583, and [O~{\sc iii}] $\lambda$5007 
echellograms of the Owl Nebula along the minor and major axes.  
The position-velocity ellipses overlaid on the echellograms represent 
the best model fit for the inner and outer shells, as described in 
\S4.1.  
The telluric H$\alpha$ line is marked on the H$\alpha$ echellograms.  
}
\end{figure}

%\clearpage

\begin{figure}
\vspace*{7.0cm}
\includegraphics{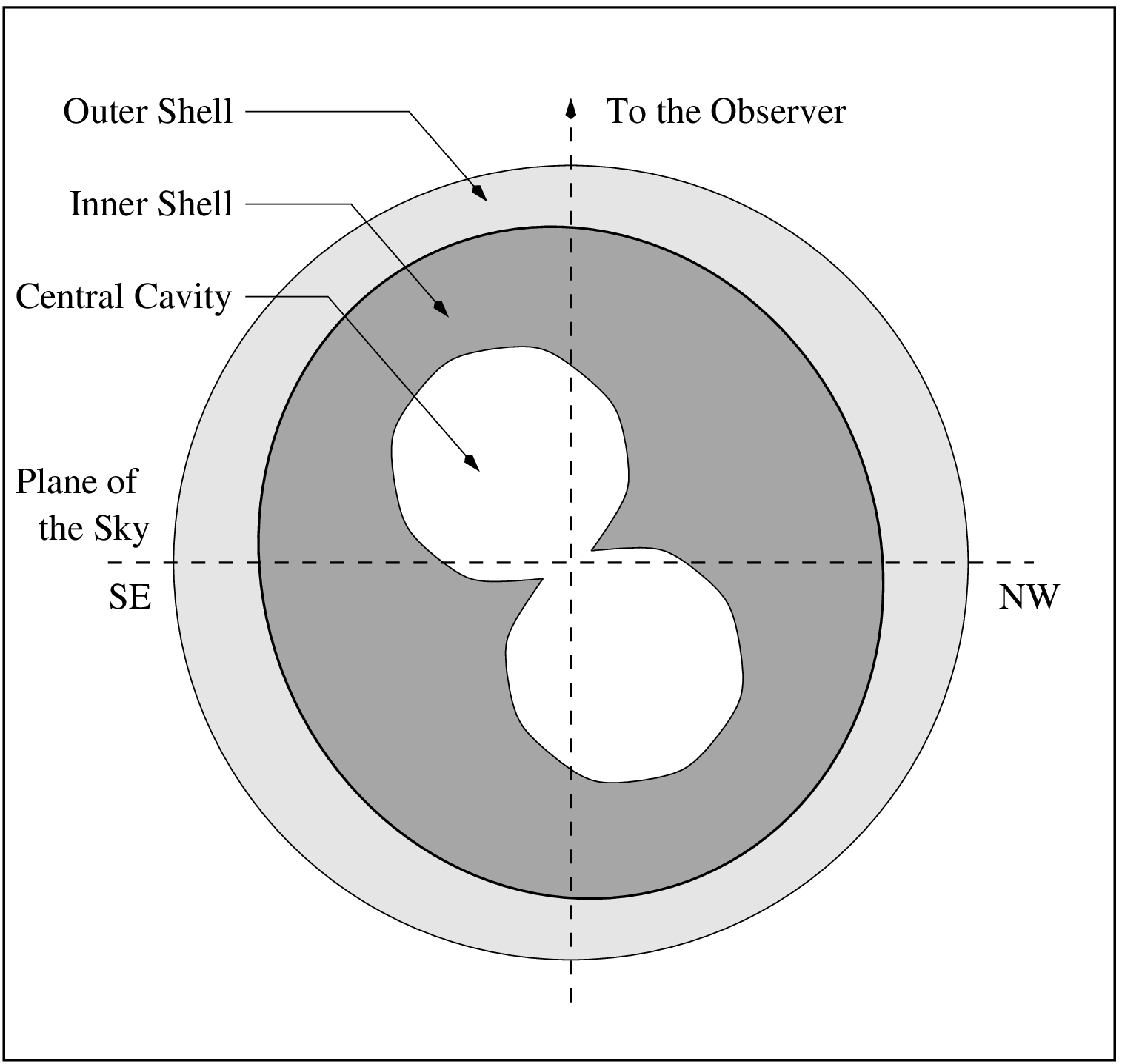}
\includegraphics{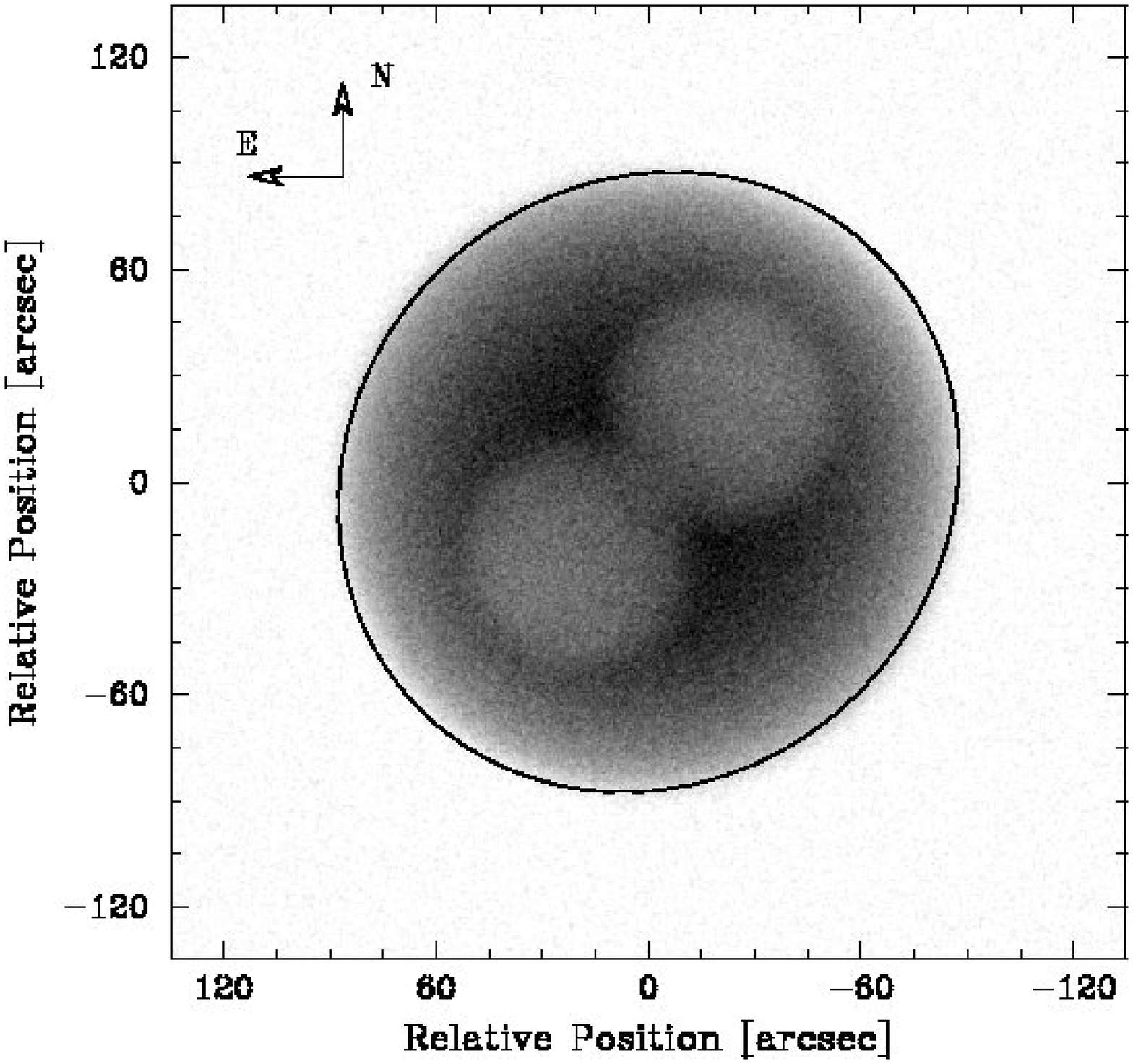}
\caption[]{
({\it Left}) 
Schematic drawing of a cross-section of our proposed model along the 
major axis of the Owl Nebula.  
The main nebula is composed of a lower-density, spherical outer shell
and an ellipsoidal inner shell with a major-to-minor axis ratio of 
$\sim$1.1.  
The inner shell has a bipolar cavity at the center, and the polar axis 
of the cavity is tilted by $\sim$30\arcdeg\ with respect to the line 
of sight.  
({\it Right}) 
Grey-scale image of the inner shell of the Owl Nebula simulated from 
our model, by projecting the inner shell shown in the left panel and 
assuming a uniform emissivity within the shell.  
The spatial resolution and signal-to-noise ratio of the simulated image 
have been selected to match those of the images.  
The ellipse marking the boundary of the inner shell shown in Fig.~1 
is overplotted on this simulated image.  
}
\end{figure}

\begin{figure}
\vspace*{6.0cm}
\includegraphics{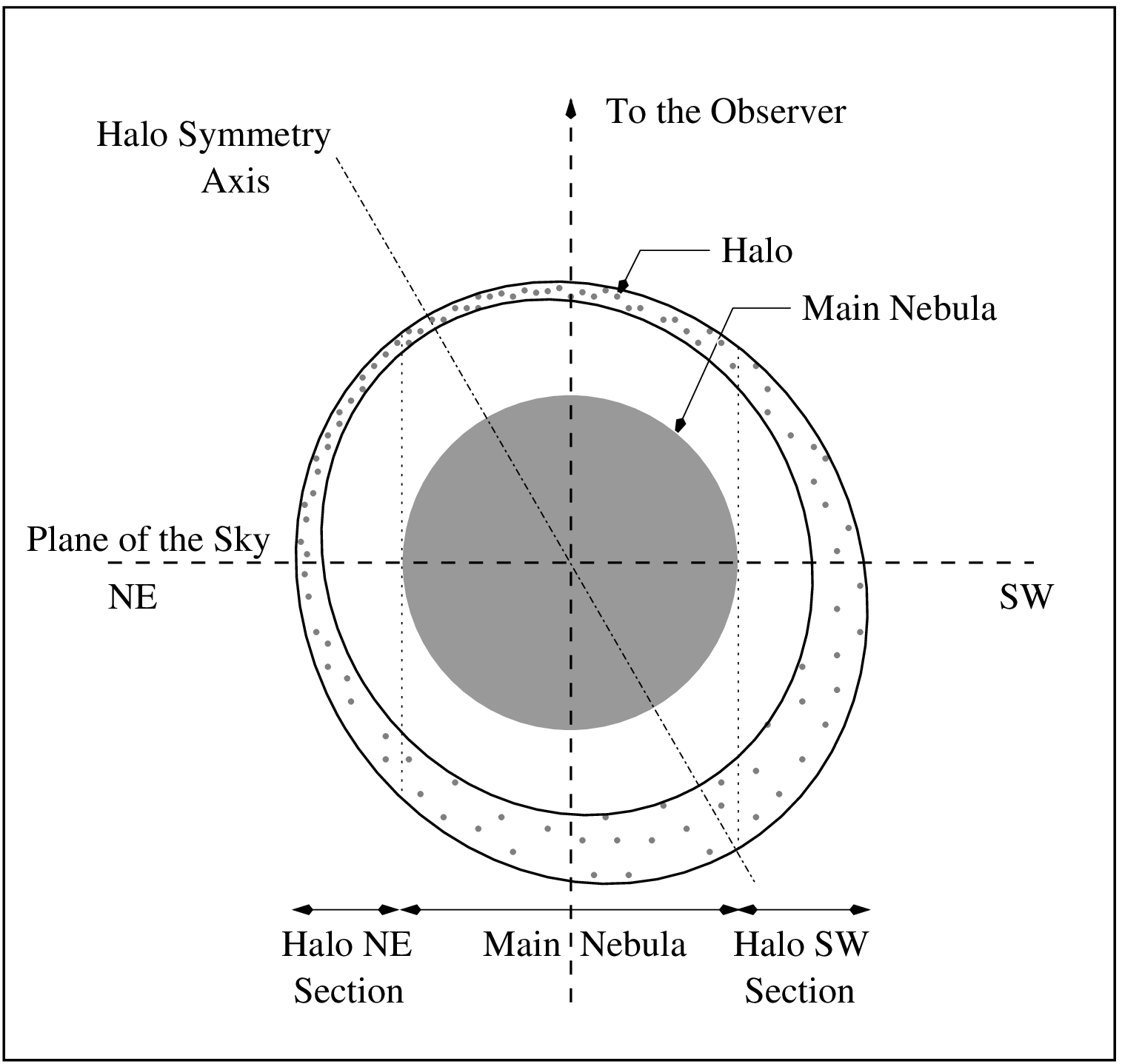}
\includegraphics{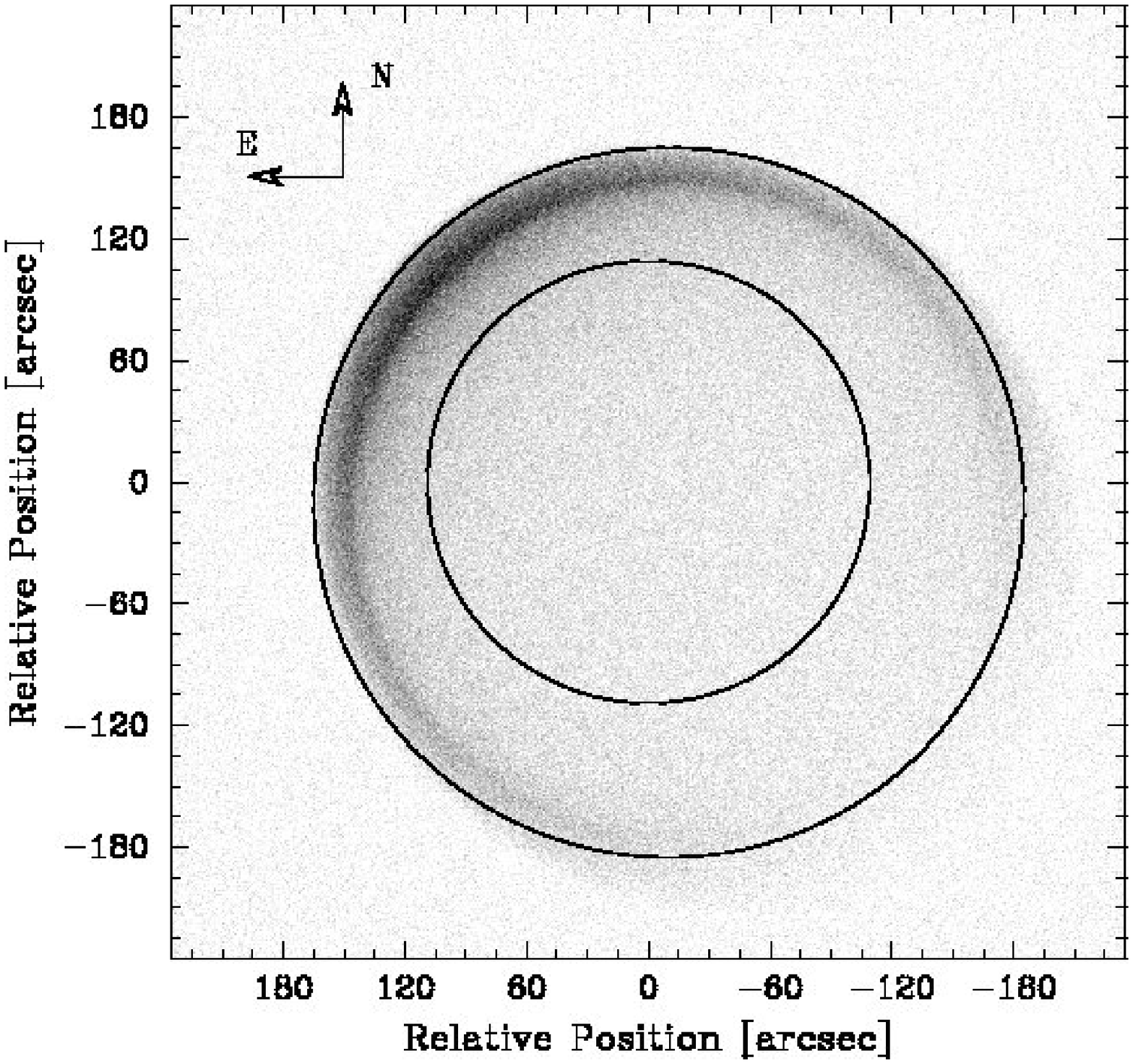}
\caption{
({\it Left}) 
Schematic drawing of a cross-section of our proposed model for the halo 
of the Owl Nebula along PA 45\arcdeg.  
The halo is shaded to indicate azimuthal variations of the volume 
emissivity.  
The filled circle at the center represents the volume occupied by the 
main nebula.  
The vertical dotted lines mark the inner edge of the halo in direct 
images.  
({\it Right}) 
Grey-scale image of the halo of the Owl Nebula simulated from 
our model, in which the halo is a 400\arcsec$\times$360\arcsec\ 
ellipsoidal shell with its major axis tilted by 30\arcdeg\ with respect 
to the line of sight.  
The contrast between the volume emissivity at the leading and trailing 
sides of the halo is 9.  
Overplotted on this image is an inner circle marking the position of 
the main nebula and an outer circle the 3 $\sigma$ contours of the 
[O~{\sc iii}] image described in Fig.~1.  
\label{sketch_halo}}
\end{figure}

\clearpage

\begin{deluxetable}{cccc}
\tablenum{1}
\tablewidth{0pt}
\tablecaption{Echelle Observations \label{tbl-1}}
\tablewidth{0pt}
\tablehead{
\multicolumn{1}{c}{Offset} & \multicolumn{1}{c}{Position Angle} & 
\multicolumn{1}{c}{Exposure Time} & \multicolumn{1}{c}{Central Wavelength} \\
\multicolumn{1}{c}{[\arcsec]} & \multicolumn{1}{c}{[\arcdeg]} & 
\multicolumn{1}{c}{[min]} & \multicolumn{1}{c}{[\AA]} 
}
\startdata

  0    & $-45$ & 30 & 6562 \\
~~~60 NW & $-45$ & 30 & 6562 \\
~150 NW & $-45$ & 30 & 6562 \\
150 SE & $-45$ & 30 & 6562 \\
  0    & $+45$ & 30 & 6562 \\
  0    & $+45$ & 30 & 5007 \\
150 NE & $+45$ & 30 & 6562 \\
150 NE & $+45$ & 30 & 6562 \\
150 SW & $+45$ & 30 & 6562 \\

\enddata
\end{deluxetable}
%% Tables should be submitted one per page, so put a \clearpage before

%% Two options are available to the author for producing tables:  the
%% deluxetable environment provided by the AASTeX package or the LaTeX
%% table environment.  Use of deluxetable is preferred.
%%

%% Three table samples follow, two marked up in the deluxetable environment,
%% one marked up as a LaTeX table.

%% In this first example, note that the \footnotesize command has been
%% used to shrink the table so it will fit on one page. Note also that
%% the \label command needs to be placed inside the \tablecaption.

\end{document}